\documentclass[
    ,final            % use final for the camera ready runs
    ,numberedheadings % uncomment this option for numbered sections
  ]
  {aipproc}

\layoutstyle{6x9}

\newcommand{\lsun}{\mbox{$L_{\odot}$}}
\newcommand{\rsun}{\mbox{$R_{\odot}$}}
\newcommand{\zsun}{\mbox{$z_{\odot}$}}
\newcommand{\Teff}{\mbox{$T_{\rm eff}$}}
\newcommand{\vinf}{\mbox{$v_{\infty}$}}

\newcommand{\Mdot}{\mbox{$\dot{M}$}}

\newcommand{\ratio}{\mbox{$v_{\infty}$/$v_{\rm esc}$}}

\newcommand{\msunyr}{\mbox{$M_{\odot} {\rm yr}^{-1}$}}
\newcommand{\Mdu}{\mbox{$\cdot 10^{-6}\,M_{\odot} {\rm yr}^{-1}$}}

\newcommand{\beq}{\begin{equation}}
\newcommand{\eeq}{\end{equation}}
\newcommand{\beqa}{\begin{eqnarray}}
\newcommand{\eeqa}{\end{eqnarray}}

\newcommand{\kms}{\mbox{${\rm km}\,{\rm s}^{-1}$}}

\newcommand{\half}{\mbox{$\frac{1}{2}$}}

\newcommand{\dd}{{\rm d}}

\newcommand{\CIII}{C\,{\sc iii}}
\newcommand{\CIV}{C\,{\sc iv}}
\newcommand{\OVI}{O\,{\sc vi}}
\newcommand{\PV}{P\,{\sc v}}

\newcommand{\FeIII}{Fe\,{\sc iii}}
\newcommand{\FeIV}{Fe\,{\sc iv}}

\newcommand{\Ha} {H$_{\rm \alpha}$}

\newcommand{\Bra}{Br$_{\rm \alpha}$}

\newcommand{\Rstar}{\mbox{$R_{\ast}$}}

\newcommand{\vesc}{\mbox{$v_{\rm esc}$}}

\newcommand{\dvdr}{\mbox{$\dd v/\dd r$}}
\newcommand{\fcl}{\mbox{$f_{\rm cl}$}}

\newcommand{\grad}{\ensuremath{g_{\rm rad}}}

\newcommand{\Neff}{\ensuremath{N_{\rm eff}}}

\newcommand{\ga}{\mathrel{\mathchoice {\vcenter{\offinterlineskip\halign{\hfil
$\displaystyle##$\hfil\cr>\cr\sim\cr}}}
{\vcenter{\offinterlineskip\halign{\hfil$\textstyle##$\hfil\cr
>\cr\sim\cr}}}
{\vcenter{\offinterlineskip\halign{\hfil$\scriptstyle##$\hfil\cr
>\cr\sim\cr}}}
{\vcenter{\offinterlineskip\halign{\hfil$\scriptscriptstyle##$\hfil\cr
>\cr\sim\cr}}}}}

\newcommand\SiIV{Si\,{\sc iv}}

\begin{document}

\title{Mass loss from OB-stars}

\classification{97.10.Me}
\keywords      {Mass loss and stellar winds}

\author{J. Puls}{
  address={Universit\"atssternwarte, Scheinerstr. 1, 81679 M\"unchen,
Germany}
}

\author{J. O. Sundqvist}{
  address={Universit\"atssternwarte, Scheinerstr. 1, 81679 M\"unchen,
Germany}
}

\author{F. Najarro}{
  address={Departamento de Astrof\'{\i}sica,
  Centro de Astrobiolog\'{\i}a, CSIC-INTA,\\
  Ctra de Torrej\'on a Ajalvir km~4,
  28850 Torrej\'on de Ardoz, Madrid,
  Spain}
}

\author{M. M. Hanson}{
  address={Department of Physics, The University of Cincinnati, Cincinnati,
OH 45221-0011}
}

\begin{abstract}
We review recent developments regarding radiation driven mass loss from
OB-stars. We first summarize the fundamental theoretical predictions, 
and then compare these to observational results (including the VLT-FLAMES 
survey of massive stars). Especially we focus on the mass loss-metallicity
dependence and on the so-called bi-stability jump.

Subsequently we concentrate on two urgent problems, weak winds and 
wind clumping, that have been identified from various diagnostics and that
challenge our present understanding of radiation driven winds. We discuss
the problems of ``measuring'' mass-loss rates from weak winds and the
potential of the near infrared, Br-alpha line as a tool to enable a more precise
quantification, and comment on physical explanations for mass-loss rates
that are much lower than predicted by the standard model.  

Wind clumping, conventionally interpreted as the consequence of a strong
instability inherent to radiative line-driving, has severe implications for
the interpretation of observational diagnostics, since derived mass-loss
rates are usually overestimated when clumping is present but ignored in the
analyses. Simplified techniques to account for clumping indicate
overestimates by factors of 2 to 10, or even more.  If actually true, these
results would have a dramatic impact on the evolution of, and the feedback
from, massive stars.  We discuss ongoing attempts (including own work)
to interprete the corresponding observations in terms of more sophisticated
models. By allowing for porosity in density and velocity space, and for a
non-void inter-clump medium, such models might require only moderate
reductions of mass-loss rates. 

\end{abstract}

\maketitle

%%%%%%%%%%%%%%%%%%%%%%%%%%%%%%%%%%%%%%%%%%%%
%% MAINMATTER
%%%%%%%%%%%%%%%%%%%%%%%%%%%%%%%%%%%%%%%%%%%%

\section{Introduction}

Massive stars are critical agents in galactic evolution, both in the present
and in the early Universe (e.g., re-ionization and first enrichment).  {\it
Mass loss} is a key process, which modifies chemical profiles, surface
abundances, and luminosities. Furthermore, {\it mass loss} has to be understood
{\it quantitatively} in order to describe and predict massive star evolution
in a correct way. The standard theory to describe hot, massive star winds is
based on radiative line-driving, and has been proven to work successfully in
most evolutionary phases (OB-stars, A-supergiants, and LBVs in their
``quiet'' phase). Also for the pivotal Wolf-Rayet (WR) stadium, line-driving
is still the most promising acceleration mechanism \citep{Graf05, Graf08}. 

In this review, we summarize fundamental predictions of the theory, as well as
corresponding observational evidence, and subsequently concentrate on two
urgent problems that challenge our understanding of line-driven winds, the
so-called weak-wind problem and wind clumping.  We concentrate on the winds
from ``normal'' OB-stars in all evolutionary phases (for corresponding
results and problems regarding WR-winds and additional material, see the
contributions by Hamann and Hillier, this volume).

\section{Line-driven winds from hot stars -- theoretical predictions}

To be efficient, radiative line-driving requires a large number of photons,
i.e., a high luminosity $L$. Since $L \propto \Teff^4 \Rstar^2$, not only
OB-supergiants, but also hot dwarfs and A-supergiants undergo significant
mass loss via this mechanism.  Typical mass-loss rates are of the order of
\Mdot\ $\approx$ 0.1 {\ldots} 10 \Mdu, with terminal velocities \vinf\
$\approx$ 200 {\ldots} 3,000 \kms. Another prerequisite is the presence of a
multitude of spectral lines, with high interaction probabilities, close to
flux maximum, implying that the strength of line-driven winds should
strongly depend on metallicity.

Pioneering work on this subject was performed by \citet{LS70} and Castor,
Abbott \& Klein (\citep{CAK}, ``CAK''), where the latter still builds the
theoretical foundation of our present understanding. Improvements with
respect to a {\it quantitative} description and first applications were
provided by \citet{FA86} and \citet{PPK}, whereas recent reviews on the
topic have been published by \citet{KP00} and \citet{Pulsrev08}.

\medskip
\noindent
{\it The principle idea of radiative line-driving} relies on two processes. \\
1. Momentum is transferred to the wind matter via line absorption/emission
processes, mostly resonance scattering, with a net change in {\it radial}
momentum
\beq 
\Delta P_{\rm radial} = \frac{h}{c}\, (\nu_{\rm in} \cos \theta_{\rm in} -
\nu_{\rm out} \cos \theta_{\rm out}) 
\eeq
where $\nu_{\rm in}$ and $\nu_{\rm out}$ are the frequencies of the absorbed
and emitted photons, and $\theta$ is the angle between the
photon's direction and the radial unit vector. 
%(parallel to the velocity vector) of the ion. 
Thanks to the fore-aft symmetry of the emission process, on average $\langle
\cos \theta_{\rm out} \rangle = 0$, whereas $\langle \cos \theta_{\rm in}
\rangle \approx 1$, since (most) of the absorbed photons originate from the
stellar surface. Thus, $\langle \Delta P_{\rm radial}\rangle \approx h
\nu_{\rm in}/c$, and the total radiative acceleration exerted on a mass
element $\Delta m$ per time interval $\Delta t$ can be derived from
considering all participating lines,
\beq
\grad = \frac{\langle \Delta P \rangle_{\rm tot}}{\Delta t \Delta m} =
\frac{\sum\limits_{{\rm all\, lines}} \langle \Delta P \rangle_{\rm i}}{\Delta t \Delta m}.
\label{grad}
\eeq

\smallskip \noindent 2. Due to the huge number of metallic lines as compared
to the few dozens from hydrogen and helium, mostly just the metal ions are
{\it directly} accelerated. Their momentum needs to be transferred to the
bulk plasma (H, He), via Coulomb collisions. The velocity drift of the metal
ions with respect to H/He is compensated for by a frictional force (``Stokes
law'') as long as the ratio between drift and thermal velocity is small
(e.g., \citep{SP92, KK00, OP02}). Otherwise (at very low wind-densities) the
metallic ions might decouple from the wind, and the wind no longer becomes
accelerated.

\smallskip \noindent The real challenge is to evaluate Eq.~\ref{grad} (see
also Owocki and Gayley, this volume). Following CAK, this is conventionally
done by (i) applying the Sobolev theory \citep{Sobo60} to approximate the
line optical depths and thus the interaction probabilities, and (ii) to
replace the summation by appropriate integrals over the line-strength 
distribution (resulting from detailed NLTE calculations), where the
line-strength $k$ is the line-opacity measured in units of the
Thomson-scattering opacity. This distribution can be fairly well
approximated by a power-law, $\dd N(k)/\dd k \propto \Neff\ k^{\alpha-2}$,
with \Neff\ the effective (flux-weighted) number of lines and $\alpha
\approx$ 0.6{\ldots}0.7 (e.g., \citep{Puls00}). Note that both quantities
depend on metallicity and spectral type. As a final result, $\grad \propto
((\dvdr)/\rho)^{\alpha}$, i.e., depends on the {\it spatial} velocity gradient
and on the inverse of the density. 

\medskip \noindent {\it Scaling relations and WLR.} Once the above
quantities are inserted into the hydrodynamic equations (adopting
stationarity), the latter can be solved (almost) analytically, returning the
following scaling relations for mass-loss rate, velocity law, and terminal
velocity:
\beqa
\Mdot &\propto& \Neff^{1/\alpha'}L^{1/\alpha'}
\Bigl(M(1-\Gamma)\Bigr)^{1-{1/\alpha'}},\qquad
v(r) = \vinf \Bigr(1-\frac{\Rstar}{r}\Bigr)^{\beta} \nonumber \\
\vinf & \approx & 2.25 \frac{\alpha}{1-\alpha}\, \vesc, \qquad \vesc=
\Bigr(\frac{2GM(1-\Gamma)}{\Rstar}\Bigr)^{\half}. 
\eeqa with Eddington-$\Gamma$, (photospheric) escape velocity $\vesc$, and
$\alpha' = \alpha - \delta$, where $\delta \approx 0.1$ describes the run of
the ionization \citep{Abbott82}. The velocity-field exponent, $\beta$, is of
the order of 0.8 (for O-stars) to 2 (for BA-supergiants).  

Using these scaling relations, a fundamental prediction for 
line-driven winds becomes apparent if one calculates the so-called 
modified wind-momentum rate,
\beq
\Mdot \vinf (\Rstar/\rsun)^{1/2} \propto 
\Neff^{1/\alpha'}L^{1/\alpha'} \Bigl(M(1-\Gamma)\Bigr)^{3/2-{1/\alpha'}},
\eeq
and accounts for the fact that $\alpha'$ is of the order of 2/3. Then the
wind-momentum rate becomes independent on mass and $\Gamma$, and can be
expressed in terms of the {\it wind-momentum luminosity relation} (WLR),
discovered first by \citet{Kud95},
\beq
\log \Bigl(\Mdot \vinf (\Rstar/\rsun)^{1/2}) \approx x \log(L/\lsun) + D(z,
\mbox{spectral type})
\label{wlr}
\eeq
with slope $x=1/\alpha'$ and offset $D$, which depends on $\Neff$ and thus
on metallicity $z$ and spectral type. Originally, it was proposed to exploit
the WLR for measuring extragalactic distances on intermediate scales (up to
the Virgo cluster), but nowadays the relation is mostly used to test the
theory itself (see below).

\medskip \noindent {\it Theoretical 1-D models.} Though the basic scaling
relations for line-driven winds are known since the key paper by CAK (and
updates by \citep{Abbott82, FA86, PPK}), {\it quantitative} predictions
require consistent NLTE/radiative-transfer calculations, to derive the
line-force as a function of spectral type and metallicity, as well as the
inclusion of processes neglected in the original work, for example
line-overlap (e.g., \citep{FC83, Puls87}). 

The most frequently cited theoretical wind models (stationary, 1-D,
homogeneous) are those from \citet{Vink00, Vink01}. Based on the 
Monte-Carlo approach developed by \citet{AL85}, they allow multi-line effects
to be considered.  In these models, the mass-loss rate is derived (iterated)
from {\it global} energy conservation, whilst the ($\beta$-) velocity field
is pre-described and the NLTE rate equations are treated in a simplified way.
\citet{Pauldrach87} and \citet{Pauldrach94, Pauldrach01}, on the other hand,
obtain a consistent hydrodynamic solution by integrating the (modified) CAK
equations based on a rigorous NLTE line-force using Sobolev line transfer.
Moreover, \citet{KK00, KK01, KK04, KK09} and \citet{Krticka06} solve the
equation of motion by means of a NLTE, Sobolev line-force, including a
more-component description of the fluid (accelerated metal ions plus H/He)
that allows them to consider questions regarding drift-velocities,
non-thermal heating, and ion decoupling.
Also, \citet{Kud02} (see also \citet{Kud89}) provides an analytic ``cooking
recipe'' for mass-loss rate and terminal velocity, based on an approximate
NLTE treatment, and
\citet{Graf05, Graf08} obtain self-consistent solutions (applied to WR
winds) by means of a NLTE line-force evaluated in the comoving frame (see
\citet{Mih75, Mih76a, Mih76b, Mih76c} and Hamann, this volume). 
Finally, \citet{Lucy07a, Lucy07b} and \citet{MV08} derive the
wind-properties from a {\it regularity} condition at the {\it sonic} point,
in contrast to most other solutions that invoke a {\it singularity}
condition at the CAK-{\it critical} point of the wind (see comments by
Owocki, this volume).

\medskip \noindent {\it Results and predictions from hydrodynamic
modeling.} Most of the various approaches yield consistent results, e.g.,
when comparing the ``mass-loss recipe'' from \citet{Vink00} with similar
investigations utilizing different codes \citep{Kud02, Pauldrach01, KK04}.
Moreover, the WLR concept is impressively confirmed by the simulations
performed by Vink et al.: The obtained modified wind-momenta follow an
almost perfect power-law with respect to stellar luminosity alone, {\it
independent of luminosity class}, and, for solar abundances, ``only'' two
distinct relations covering the complete spectral range have been found, one
for 50~kK $>$ \Teff\ $>$ 27.5~kK and the other for 22.5~kK $>$ \Teff\ $>$
12~kK, respectively. In other words, the spectral type dependence of $x$ and
$D$ in Eq.~\ref{wlr} seems to be rather mild.

Also regarding the predicted metallicity dependence, the various results
agree satisfactorily (note that the $z$-dependence of \vinf\ is rather 
weak):
\begin{eqnarray*}
\mbox{\citet{Kud02}:} \qquad \vinf  &\propto& z^{0.12},
\qquad \mbox{\citet{Krticka06}:} \qquad \vinf \, \propto \, z^{0.06}, \\
\mbox{\citet{Vink01}:}\qquad  \Mdot &\propto& z^{0.69} \mbox{  for
O-stars}, \qquad \Mdot \, \propto \, z^{0.64} \mbox{  for B-supergiants},\\
\mbox{\citet{Krticka06}:}\qquad  \Mdot &\propto& z^{0.67} \mbox{  for
O-stars.} 
\end{eqnarray*}

\section{Observations vs. Theory}

In the last decade, various spectroscopic NLTE analyses of hot stars {\it
and their winds} have been undertaken, in the Galaxy and in the Magellanic
Clouds, in the UV, in the optical, and in a combination of both.  For a
compilation of these publications (without Galactic Center objects and
objects analyzed within the {\sc flames} survey of massive stars, see
below), see Tables 2 and 3 in \citet{Puls08}, to be augmented by the
UV-P{\sc v} investigation of Galactic O-stars by \citet{Fullerton06}, the
UV+optical analysis of Galactic O-dwarfs by \citet{Marcolino09}, and the
optical analysis of LMC/SMC O-stars by \citet{Massey09}. Most of this work
has been performed by means of 1-D, line-blanketed, NLTE,
atmosphere/spectrum-synthesis codes allowing for the presence of winds, in
particular {\sc cmfgen} (\citet{HillierMiller98}), {\sc wm-b}asic
(\citet{Pauldrach01}), and {\sc fastwind} (\citet{Puls05}).

\medskip \noindent {\it Central results.} The results of these
investigations can be roughly summarized as follows. (i) The mass-loss rates
from SMC stars (with $z \approx 0.2\ \zsun$, see \citep{Mokiem07b} and
references therein) are indeed lower than those from their Galactic
counterparts. (ii) For O- and early B-stars, the theoretically predicted WLR
from \citet{Vink00} is met, except for O-supergiants with rather dense
winds, in which the observed wind-momenta are higher (by factors around
three) than the predictions (which might be explained by wind-clumping
effects, see Sect.~\ref{clumping}), and for a number of late O-dwarfs (and a
few O-giants), in which the observed wind-momenta are much lower than the
predictions (this is the so-called ``weak-wind problem'', see 
Sect.~\ref{weakwinds}). (iii) B-supergiants below the ``bi-stability jump''
(\Teff $<$ 22~kK) show lower wind-momenta than predicted, as outlined in the
following.

\medskip \noindent {\it The bi-stability jump: predictions and
observations.} A fundamental prediction by \citet{Vink00} is the occurrence of
two distinct WLRs, one for hotter objects and one for cooler objects, with
the division located around 25$\pm$2.5~kK. This rather abrupt change is due
to the so-called bi-stability mechanism\footnote{denoted after some peculiar
behaviour of theoretical models for the wind of P~Cygni \citep{PP90}.},
which relies on the fact that the mass-loss rates of line-driven winds are,
for typical chemical compositions, primarily controlled by the number and
distribution of {\it iron-}lines, because of their dominant contribution
($\sim$50\%) to the total line acceleration in the lower wind \citep{Puls00,
Vink00, Krticka06}. Below roughly 25~kK, the ionization of iron is predicted
to switch abruptly from \FeIV\ to \FeIII, and since \FeIII\ has more driving
lines than \FeIV\ at flux maximum, the mass-loss rate must increase.
Quantitatively, \citet{Vink00} predict an increase in \Mdot\ by a factor of
five and a decrease of \vinf\ by a factor of two, so that, overall,
B-supergiants (except for the earliest sub-types) should have higher
wind-momenta than their O-star counterparts at the same luminosity.

Observations confirm the ``velocity-part'' of this picture, at least
qualitatively. For stars with \Teff$\ge$ 23~kK, the observed ratio is 
\ratio\ $\approx$~3, whereas it decreases {\it gradually} towards cooler
temperatures, reaching values of \ratio$\approx$ 1.3{\ldots}1.5 for stars
with \Teff$\le$ 18~kK \citep{Evans04, Crowther06, MP08}. With respect to the
predicted increase in \Mdot, however, the situation is different. As shown
by \citet{MP08}, the mass-loss rates of B-supergiants below the {\it
observed} bi-stability jump (\Teff $<$ 22~kK) actually {\it decrease} or at
least do no change. This is a first indication that there are still problems
in our understanding of line-driven winds.

\medskip \noindent {\it The {\sc flames} survey of massive stars.} Further
progress has been obtained within the {\sc flames} survey of massive stars
(P.I. S. Smartt), a project that performed high resolution multi-object
spectroscopy of stars located within eight young and old clusters in the
Galaxy and the Magellanic Clouds. In total, 86 O-stars and 615 B-stars were
observed (for introductory papers and a brief summary, see \citet{Evans05,
Evans06, Evans08}). The major scientific objectives of this survey were to
investigate (i) the relation between stellar rotation and abundances (i.e.,
to test the present theory of rotational mixing), (ii) the role of binarity,
and (iii) stellar mass-loss as a function of metallicity.

Regarding the last objective, \citet{Mokiem06, Mokiem07a} analyzed a total
of $\sim 60$ O- and early B-stars in the SMC and LMC, by means of {\sc
fastwind} and using a genetic algorithm \citep{Mokiem05}. The results were
combined by \citet{Mokiem07b} with data from previous investigations, to
infer the metallicity dependence of line-driven mass-loss based on a
significant sample of stars.  Using mean abundances of $z=0.5\ \zsun$ (LMC)
and $z=0.2\ \zsun$ (SMC), a metallicity dependence of $\vinf \propto
(z/\zsun)^{0.13}$, and a correction for clumping effects (see below)
following \citet{Repo04}, they derived an {\it empirical} relation 
\beq 
\Mdot \propto (z/\zsun)^{0.72\pm0.15}, 
\eeq 
with rather narrow confidence intervals. This result is consistent with
theoretical predictions, both from line-statistics \citep{Puls00} and from
hydrodynamic models (see above). 

\section{Weak winds}
\label{weakwinds}

The results as summarized above imply that line-driven mass loss seems to be
basically understood, though certain problems need further consideration. 
In particular, from early on there were indications that the (simple) theory
might break down for low-density winds. E.g., \citet{Chlebo91} have derived
mass-loss rates for late O-dwarfs that are factors of ten lower than
expected.  By means of UV-line diagnostics, \citet{KudEUV91} and
\citet{Drew94} have derived mass-loss rates for two BII stars that are a
factor of five lower than predicted, and \citet{Puls96} have shown that the
wind-momentum rates for low-luminosity dwarfs and giants ($\log L/L_{\odot} 
<$ 5.3) lie well below the empirical relation for ``normal'' O-stars. 

The last investigation illuminated an immediate problem arising for
low-density winds. For $\Mdot < (5{\ldots}1) \cdot 10^{-8}$~\msunyr, the
conventional mass-loss indicator, \Ha, becomes insensitive, and only upper
limits for \Mdot\ can be derived (for a recent illustration of this problem,
see \citep{Marcolino09}). Instead, unsaturated UV resonance lines (\CIV,
\SiIV, \CIII) might be used to obtain actual values for \Mdot\ (e.g.,
\citep{Martins04, Pulsrev08, Marcolino09}). 

By means of such UV-diagnostics, strong evidence has accumulated that a
large number of late type O-dwarfs (and a few giants of intermediate
spectral type) have mass-loss rates that are factors of 10 to 100 lower than
corresponding rates from both predictions and extrapolations of empirical
WLRs. In particular, such {\it weak winds} have been found in the Magellanic
Clouds (O-dwarfs in NCG\,346 (LMC): \citet{Bouret03}; extremely young
O-dwarfs in N81 (SMC): \citet{Martins04}) and in the Milky Way (O-dwarfs and
giants: \citet{Martins05}; late O-dwarfs: \citet{Marcolino09}). 

Two points have to be stressed. (i) Until now, it is not clear whether {\it
all} or only part of the late type dwarfs are affected by this problem. (ii)
The derived UV mass-loss rates are not very well constrained, since they
might be contaminated\footnote{via a modified ionization equilibrium.} from
X-rays embedded in the wind (due to shocks, see next Section). The higher
the X-ray emission, the weaker the lines, and the higher the {\it actual} 
mass-loss rates (see Figs. 19 and 20 in \citet{Pulsrev08}). However, to
``unify'' the present, very low, \Mdot-values with ``normal'' mass-loss
rates by invoking X-rays, {\it unrealistically high} X-ray luminosities
would be required \citep{Marcolino09}. 

The weak-wind problem is a prime challenge for the radiative line-driven
wind theory. \citet{Martins04} investigated a variety of candidate processes
(e.g., ionic decoupling, shadowing by photospheric lines, curvature effects
of velocity fields), but none of those turned out to be strong enough to
explain the very low mass-loss rates that seem to be present. At the end of
this review, we will return to this problem.

\section{Wind clumping}
\label{clumping}

During the last years, overwhelming direct and indirect evidence has
accumulated that one of the standard assumptions of conventional wind
models, {\it homogeneity}, needs to be relaxed. Nowadays the winds are
thought to be clumpy, consisting of {\it small scale} density
inhomogeneities, where the wind matter is compressed into over-dense clumps,
separated by an (almost) void inter-clump medium ({\sc icm}). Details on
observations and theory can be found in the proceedings of a recent 
workshop, ``Clumping in hot star winds'' \citep{Hamann08}. 

Theoretically, such inhomogeneities are considered related to structure
formation due to the line-driven (``de-shadowing'') instability, a strong
instability inherent to radiative line-driving (cf. Owocki, this volume).
Time-dependent hydrodynamic models allowing for this instability to operate
have been developed by Owocki and coworkers (1-D:\citep{OCR88, RO02, RO05};
2-D:\citep{DO03, DO05}) and by Feldmeier \citep{Feldmeier95, Feldmeier97},
and show that the wind, for $r \ga 1.3 \Rstar$, develops extensive structure
consisting of strong {\it reverse} shocks separating slower, dense material
from high-speed rarefied regions in between. Such structure is the most
prominent and robust result from time-dependent modeling, and {\it the
basis for our interpretation and description of wind clumping}. Within the
shocks, the material is heated to a couple of million Kelvin, and
subsequently cooled by X-ray emission (which has been observed by all X-ray
observatories), with typical X-ray luminosities $L_{\rm X}/L_{\rm bol}
\approx 10^{-7}$ (for newest results, see \citet{Sana06}).

\medskip \noindent {\it Clumping effects.} Until now, most diagnostic
methods to investigate the effects of clumping use the following
assumptions: The clumps are {\it optically thin}, the {\sc icm} is void, the
velocity field remains undisturbed, and the so-called clumping factor, \fcl,
measures the over-density inside the clumps with respect to the average
density. This simple model of {\it micro-clumping} allows one to incorporate
clumping into NLTE-codes without any major effort, namely by multiplying the
average (wind-) density by \fcl\ and by multiplying all
opacities/emissivities by the inverse of \fcl\ (i.e., by the volume filling
factor).

The most important consequence of such optically thin clumps is a reduction
of any \Mdot\ derived from $\rho^2$-dependent diagnostics (e.g.,
recombination based processes such as \Ha\ or radio-emission), assuming
smooth models, by a factor of $\sqrt f_{\rm cl}$. That there is a {\it
reduction} is conceivable, since, under the assumptions made, the square of
the over-density ``wins'' against the smaller absorbing/emitting volume.
Thus, a lower \Mdot\ is sufficient to produce the same optical
depths/emission measures as in smooth models.

Note, however, that in this scenario any \Mdot\ derived from
$\rho$-dependent diagnostics (e.g., UV-resonance lines) remains
uncontaminated, since in this case the over-density cancels against the
smaller absorbing/emitting volume. Finally, it should be mentioned that a
clumpy medium also affects the ionization equilibrium, due to enhanced
recombination (e.g., \citep{Bouret05}).

Results from NLTE-spectroscopy allowing for micro-clumped winds are as
follows. (i) Typical clumping factors are \fcl $\approx$ 10{\ldots}50, and
clumping starts at or close to the wind base, the latter in conflict with 
theoretical predictions. Derived mass-loss rates are factors of 3 to 7 lower
than previously thought \citep{Crowther02, Hillier03, Bouret03, Bouret05}.
In strong winds, the inner region is more clumped than the outer one
($f_{\rm cl}^{\rm in} \approx 4{\ldots}6 \times f_{\rm cl}^{\rm out}$), and
the minimum reduction of smooth \Ha\ mass-loss rates is by factors between 2
and 3 \citep{Puls06}.

\medskip \noindent {\it The \PV\ problem.} From a mass-loss analysis using 
the FUV \PV\ resonance line\footnote{unsaturated due to the low phosphorus
abundance.} for a large sample of O-stars, \citet{Fullerton06} (see also
\citep{Massa03}) concluded that the resulting mass-loss rates are {\it a
factor of 10 or more lower} than derived from \Ha\ and/or radio emission
using homogeneous models, implying \fcl $\ga$ 100! Similar results have been
found from unsaturated P~Cygni lines from lower luminosity B-supergiants
\citep{Prinja05}.

If such large reductions in \Mdot\ were true, the consequences for stellar
evolution and feed-back would be enormous. Note that an ``allowed''
reduction from evolutionary constraints is at most by a factor of 2 to 4
(\citet{Hirschi08}).

\medskip \noindent {\it Porosity and vorosity.} A possible resolution of
this dilemma might be provided by considering the {\it porosity}
\citep{Owocki04} of the medium, also suggested to explain the observed X-ray
line emission (cf. the contributions by Oskinova et al. and Cohen et al. in
\citep{Hamann08}, and particularly the discussion on X-rays). Whenever the
clumps become optically thick for certain processes, as might be true for
the \PV-line, the geometrical distribution of the clumps becomes important
(size vs. separation, shape). In this {\it macro-clumping} approach (see
also Hamann, this volume), the effective opacity becomes reduced, i.e., the
wind becomes more transparent (``porous''), because radiation can propagate
through the ``holes'' provided by the {\sc icm}.  Additionally, clumps hidden 
behind other clumps become ineffective because the first clump is already 
optically thick.

\citet{Oskinova07} used a simple, quasi-analytic treatment of macro-clumping
(still assuming a smooth velocity law) to investigate \PV\ in parallel with
\Ha\ from $\zeta$~Pup. Whereas macro-clumping had almost no effect on \Ha,
since the transition is optically thin in the clumps, \PV\ turned out to be
severely affected. Thus, only a moderate reduction of the smooth mass-loss
rate (factors 2 to 3) was necessary to fit the observations, consistent with
the evolutionary constraints from above. 

This model has been criticized by \citet{Owocki08}, who pointed out that not
only the distribution/optical thickness of the clumps is important, but also
the distribution of the velocity field, since the interaction between
photons and {\it lines} is controlled by the Doppler-effect. Also the
``holes'' in {\it velocity space}, due to the non-monotonic character of the
velocity field, lead to an increased escape (thus, he called this process
velocity-porosity = ``vorosity''), whilst the different velocity gradients
inside the clumps lead to an additional modification of the optical depth.

\begin{figure}
%\hspace{-0.8cm}
\begin{minipage}{8.0cm}
\resizebox{\hsize}{!}
   {\includegraphics[angle=90]{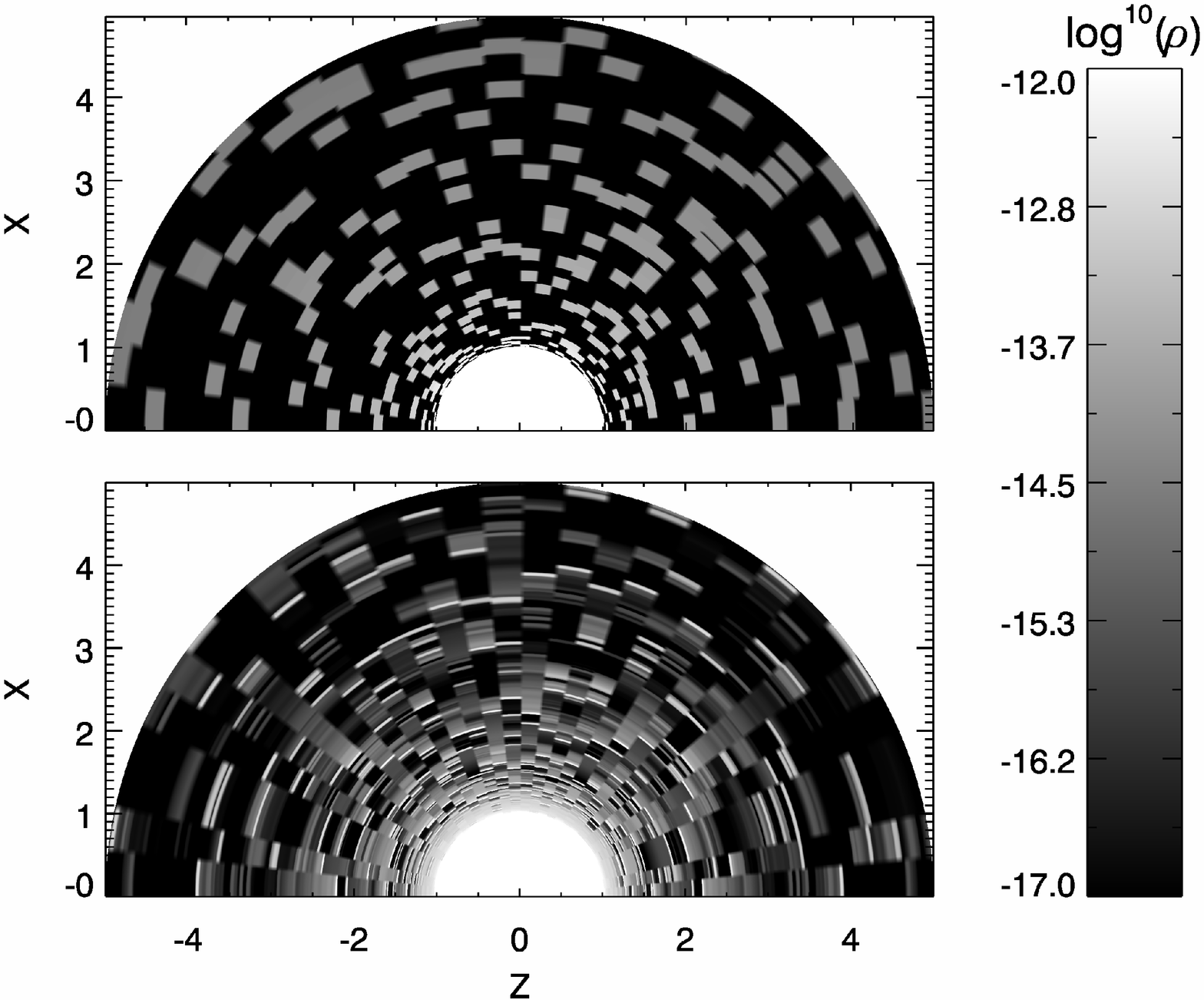}}
\end{minipage}
\hspace{-1.2cm}
\begin{minipage}{8.5cm}
   \resizebox{\hsize}{!}
   {\includegraphics[angle=90]{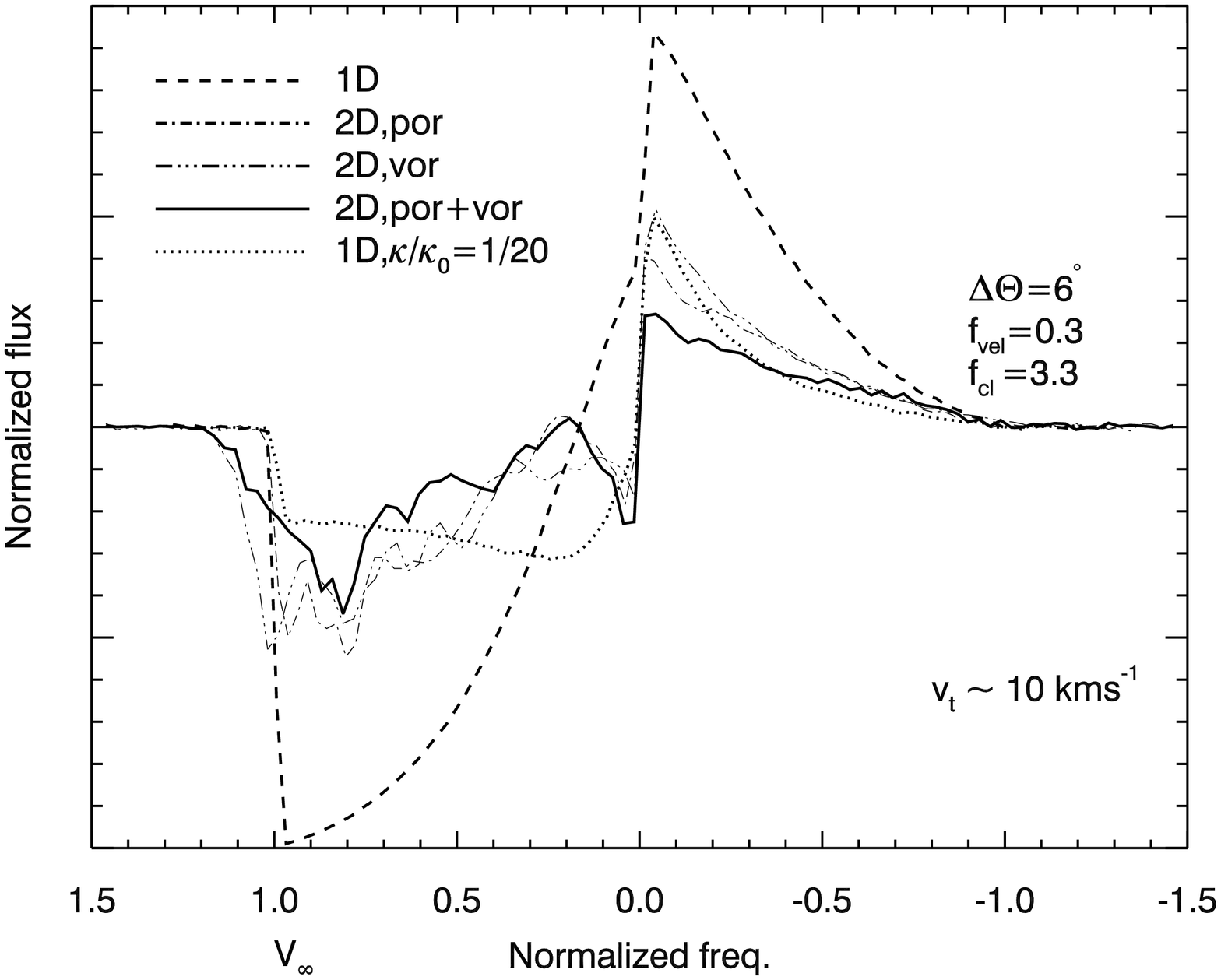}}
\end{minipage}
  \caption{{\bf Left}: Density contours of stochastic (upper) and pseudo 2-D
hydrodynamic wind models as investigated by our group. {\bf Right:} Line
profiles for an intermediate strong line formed in inhomogeneous winds with 
different sub-structures. See text.}
\label{2dwinds}
\end{figure}

\medskip \noindent {\it Resonance line formation with porosity and
vorosity.} To clarify in how far the above arguments/simulations depend on
the various assumptions, and to characterize/quantify the various effects
from inhomogeneous winds of {\it different sub-structures}, a current
project in our group investigates the resonance-line formation in such
winds. To this end, pseudo 2-D hydrodynamic models (based on different
snapshots of corresponding 1-D models from Owocki and Feldmeier, aligned as
independent slices of opening angle $\Theta$), as well as 2-D models based
on a stochastic description, have been created (Fig.~\ref{2dwinds}, left
panel).  For these models then, a {\it detailed} Monte-Carlo line transfer
(discarding the Sobolev-approximation) is performed. The right panel of
Fig.~\ref{2dwinds} shows prototypical profiles from such simulations, based
on the stochastic 2-D wind description, for an intermediate strong line that
would be marginally saturated in smooth models (dashed). The grey
dashed-dotted profile displays the effects of porosity alone (i.e., a smooth
velocity field has been used), using a rather low clumping factor, $\fcl =
3.3$, and an average separation of clumps $\sim
\Rstar$ in the outer wind.  Already here, a strong de-saturation of the profile is visible. The
grey dashed-dotted-dotted line displays the other extreme, namely vorosity
alone (i.e., now the density is smooth), using a stochastic description of
the velocity field, characterized by a ``velocity clumping factor'' (as
defined in \citep{Owocki08}, Fig. 1) $f_{\rm vel}$ = 0.3.  Interestingly,
the de-saturation of the profile is similar to the porosity-effect alone.
The solid black line displays the combined effect from porosity and
vorosity, with a further de-saturation. If compared to a line from a smooth
model of similar profile strength (dotted), it turns out that the effective
opacity in the structured model(s) has been reduced by a factor of 20, i.e.,
the actual \Mdot\ would be a factor of 20 higher than derived from a smooth
model. Thus, structured models invoking porosity and vorosity might indeed
resolve the discordance between the results by Fullerton et al. and
evolutionary constraints. 

We note, however, that the profile-strength reduction presented in
Fig.~\ref{2dwinds} corresponds to a ``most favourable case'', using rather
ideal parameters. Our investigations have shown how details on porosity,
vorosity, and the {\sc icm}, all are important for the formation of the line
profiles. In fact, the strengths of similar profiles calculated from our
pseudo 2-D hydrodynamic models are only reduced by $\approx 10 \%$, because
of insufficient vorosity inherent to structures from present time-dependent
modeling (see also \citet{Owocki08}). Such a
modest reduction is much lower than needed to alleviate the discrepancy
discussed above. Also, as it turns out, the {\sc icm} is a crucial parameter
if to de-saturate intermediate strong lines and, at the same time, allowing
the formation of the observed saturated profiles. Tests have shown that,
with a void {\sc icm}, the formation of saturated profiles is only possible 
if the average clump separation (controlling the porosity) is very small,
but then the de-saturation of intermediate strong lines becomes marginal.
Only by assuming an {\sc icm} with sufficient density ($\approx 0.01
\rho_{\rm smooth}$) we have been able to form saturated lines in parallel
with de-saturated ones of intermediate strength. This finding is consistent
with results from \citet{Zsargo08}, who pointed out that the {\sc icm} is
crucial for the formation of highly ionized species such as \OVI.

Further details and results from our investigations will be given in a
forthcoming paper (Sundqvist et al., in prep for A\&A), including a
systematic investigation of different key parameters and effects. Future
plans include a comparison with emission lines, and the development of
simplified approaches to incorporate porosity/vorosity effects into NLTE
models.

\section{Weak winds again -- \Bra\ as a diagnostic tool}
\label{Bra}

In the preceeding paragraphs, we have argued that (i) mass-loss rates from
unsaturated UV line-profiles are {\it much} lower than those from \Ha\ or
radio emission, and that (ii) this discordance might be mitigated by
porosity/vorosity effects. Recall here that the mass-loss rates from weak
winds discussed so far (Sect.~\ref{weakwinds}) rely on the same UV-line
diagnostics, and the question arises whether one encounters a similar
problem, i.e., an under-estimation of the ``true'' mass-loss rates due to
insufficient physics accounted for in the diagnostics. Thus, to clarify in
how far the weak wind problem is a real one, independent diagnostics are
required! 

Already in 1969 \citet{Auer69}, based on their first generation of NLTE,
hot-star model atmospheres, predicted that the IR \Bra-line should show
significant {\it photospheric} core emission, due to an under-population of
its lower level ($n=4$) relative to the upper one ($n=5$), resulting from a
very efficient decay channel $4 \rightarrow 3$. Indeed, such core emission
has meanwhile been observed in various weak wind candidates such as
$\tau$~Sco (B0.2V), HD\,36861 (O8III(f)), and HD\,37468 (O9.5V) (Najarro,
Hanson and Puls, in prep. for A\&A). Recent simulations (\citet{Pulsrev08},
Figs.~21/22) actually show that such photospheric + wind emission can fit
the observations quite nicely, and that the core of \Bra\ is a perfect
tracer for the wind density also for thinner winds (as opposed to \Ha).
Astonishingly, the {\it height of the peak increases for decreasing \Mdot},
which is related to the {\it onset} of the wind, i.e., the density/velocity
structure in the transition zone between photosphere and wind, and not due
to radiative transfer effects. The higher the wind-density, the deeper (with
respect to optical depth) this onset, which subsequently suppresses the
relative under-population of $n=4$ due to efficient pumping from the
hydrogen ground-state.  More\-over, \Bra\ is only weakly affected by the
presence of X-rays, and thus an ideal tool to infer very low mass-loss
rates. From fits to the observations, it turns out that $\Mdot$ is actually
very low (of the order of $10^{-10} \msunyr$ for HD\,37468, and even lower,
if the wind-base were clumped). {\it Thus, weak winds seem to be a reality!}

What may then be the origin of weak winds? \citet{KK09} argue that 
weak-winded stars display enhanced X-ray emission, maybe related to extended
cooling zones because of the low wind density. Already \citet{Drew94}
pointed out that strong X-ray emission can lead to a reduced line
acceleration, because of a modified ionization equilibrium, and since higher
ions have fewer lines. Thus, weak-winded stars might be the result of strong
X-ray emission. Let us now speculate whether such strong emission might be
related to magnetic fields. Note that weak winds can be strongly affected by
relatively weak $B$-fields, of the order of 40 Gauss according to the
scaling relations provided by \citet{udDoula02}, which is below the present
detection threshold. In this case then, colliding loops might be generated,
which in turn generate strong and hard X-ray emission in the lower wind,
which finally might influence the ionization and thus radiative driving.
Future simulations coupling magneto-radiation-hydrodynamic wind codes with a
{\it self-consistent} description of the line-acceleration will tell whether
this mechanism might work.

%%%%%%%%%%%%%%%%%%%%%%%%%%%%%%%%%%%%%%%%%%%%%%%%
%% BACKMATTER
%%%%%%%%%%%%%%%%%%%%%%%%%%%%%%%%%%%%%%%%%%%%%%%%

\begin{theacknowledgments}
J. Puls gratefully acknowledges a travel grant (Pu~117/7-1) by the German
DFG. Part of this work has been supported by the International
Max-Planck Research School of Astrophysics (IMPRS), via a grant to
J.~Sundqvist. F. Najarro acknowledges a Spanish AYA2008-06166-C03-02 grant. 
Contributions to this work by M.M. Hanson were supported by the National Science
Foundation Grant No. 0607497.

\end{theacknowledgments}

\bibliographystyle{aipproc}   % if natbib is available

%\bibliography{puls}

\end{document}